\documentclass[letterpaper, 10 pt, conference]{ieeeconf}  % Comment this line out if you need a4paper

\IEEEoverridecommandlockouts                              % This command is only needed if 
\usepackage{float}
\usepackage{graphicx}
\usepackage{subcaption}
\usepackage{tabularx}
\usepackage{amssymb}
\usepackage{amsmath} 
\usepackage{array}

\title{\LARGE \bf
ConvoyLLM: Dynamic Multi-Lane Convoy Control Using LLMs
}

\author{Liping Lu$^{1}$, Zhican He$^{1}$, Duanfeng Chu$^{2*}$, Rukang Wang$^{2}$, Saiqian Peng$^{2}$, Pan Zhou$^{3}$% <-this % stops a space
\thanks{*This work is supported in part by the National Natural Science Foundation of China (52472438), the Natural Science Foundation of Hubei Province for Distinguished Young Scholars (2022CFA091), the Key R\&D Program of Hubei Province (2024BAB033), and Wuhan Science and Technology Major Project (2022013702025184).}% <-this % stops a space
\thanks{$^{1}$L. Lu and Z. He are with the School of Computer Science and Artificial Intelligence, Wuhan University of Technology, Wuhan 430070, China.}%
\thanks{$^{2}$D. Chu, R. Wang and S. Peng are with Intelligent Transportation Systems Research Center, Wuhan University of Technology, Wuhan 430070, China.}%
\thanks{$^{3}$P. Zhou is with Hubei Key Laboratory of Distributed System Security, School of Cyber Science and Engineering, Huazhong University of Science and Technology, Wuhan 430074, China.}%
\thanks{$^{*}$D. Chu is the corresponding author. Email: {\tt\small chudf@whut.edu.cn}}%
}

\begin{document}

\maketitle
\thispagestyle{empty}
\pagestyle{empty}
\bibliographystyle{unsrt}

%%%%%%%%%%%%%%%%%%%%%%%%%%%%%%%%%%%%%%%%%%%%%%%%%%%%%%%%%%%%%%%%%%%%%%%%%%%%%%%%
\begin{abstract}

This paper proposes a novel method for multi-lane convoy formation control that uses large language models (LLMs) to tackle coordination challenges in dynamic highway environments. Each connected and autonomous vehicle in the convoy uses a knowledge-driven approach to make real-time adaptive decisions based on various scenarios. Our method enables vehicles to dynamically perform tasks, including obstacle avoidance, convoy joining/leaving, and escort formation switching, all while maintaining the overall convoy structure. We design a Interlaced formation control strategy based on locally dynamic distributed graphs, ensuring the convoy remains stable and flexible. We conduct extensive experiments in the SUMO simulation platform across multiple traffic scenarios, and the results demonstrate that the proposed method is effective, robust, and adaptable to dynamic environments. The code is available at: {\tt\small {https://github.com/chuduanfeng/ConvoyLLM}} .

\end{abstract}

%%%%%%%%%%%%%%%%%%%%%%%%%%%%%%%%%%%%%%%%%%%%%%%%%%%%%%%%%%%%%%%%%%%%%%%%%%%%%%%%
\section{INTRODUCTION}

With the rapid development of Connected and Automated Vehicles (CAVs) technology, convoy coordination control has shown significant potential in improving traffic flow efficiency, driving safety, and fuel economy. In multi-lane highway scenarios, through proper convoy formation control, it is possible to achieve efficient coordination between vehicles, significantly alleviate traffic congestion, and reduce energy consumption. However, most past research has focused on single-lane convoy control, primarily addressing longitudinal following behaviors \cite{salek2024overview}. There is limited research on convoy formation control in complex dynamic environments of multi-lane highways, particularly concerning how to maintain formation stability and achieve adaptive adjustments in dynamic scenarios, which remains an urgent challenge.

Formation control methods, from a global perspective, treat multiple CAVs as a whole, dynamically adjusting the convoy's formation according to changes in the driving environment. Some progress has been made in this area \cite{nahavandi2022autonomous}, with methods demonstrating the potential for dynamic formation adjustment in multi-lane scenarios. However, existing approaches still face the following limitations: (1) rule-based control strategies are struggle to handle complex decision-making in dynamic multi-lane scenarios; (2) methods such as reinforcement learning require large amounts of training data and have limited generalization ability.

In recent years, the development of large language models (LLMs) has provided new possibilities for addressing these issues. From the early statistical language models to deep learning models based on Transformer, LLMs have evolved rapidly. Recent models such as GPT-4 \cite{achiam2023gpt}, GPT-4o, and LLaMA3 \cite{dubey2024llama} have shown exceptional capabilities in semantic understanding and generation tasks. The advantage of LLMs lies in their strong reasoning ability and understanding of complex semantic relationships, enabling them to process multimodal inputs and generate outputs that meet specific contextual requirements. These characteristics make LLMs widely applicable in autonomous driving, such as for traffic situation analysis, behavioral strategy generation, and navigation and planning \cite{zhang2024advancing}.

In this paper, we introduce LLMs into multi-lane convoy control for the first time to enhance the convoy's adaptive capabilities and cooperation in complex scenarios. The main role of the LLM is to provide effective decision-making support for each vehicle in the convoy through efficient reasoning and generation capabilities. At the same time, we integrate a staggered formation control strategy based on locally dynamic distributed graphs, ensuring that the convoy maintains formation while exhibiting both flexibility and stability.

\section{RELATED WORK}

\subsection{Formation Decision Making} Formation decision-making plays a crucial role in enhancing traffic efficiency, safety, and fuel economy, particularly in the context of CAVs. Early research on convoy decision-making primarily focused on coordination in single-lane scenarios, emphasizing vehicle longitudinal behaviors such as maintaining constant speed and ensuring safe inter-vehicle distance \cite{jia2015survey,li2019platoon}. However, these methods primarily addressed simple longitudinal vehicle behaviors and lacked the adaptability required to handle the complexities of real-world driving environments.

With the development of multi-lane highways and more complex traffic dynamics, convoy decision-making algorithms need to exhibit stronger adaptability. A significant challenge is how to effectively adjust the convoy's behavior in dynamic environments, including lane changing, obstacle avoidance, and adapting to traffic congestion. For example, Gao et al. \cite{gao2019multi} used the artificial potential field method to improve obstacle avoidance, while Cai et al. \cite{cai2019multi,cai2021formation} developed dynamic staggered formation generation methods, they also introduced a dual-layer planning framework to optimize lane changes and formation adjustments in real-time, improving traffic efficiency and vehicle coordination. However, these methods often rely on predefined strategies and lack flexibility in responding to unexpected scenarios.

In recent years, data-driven deep reinforcement learning (RL) has introduced a new paradigm for convoy decision-making. Xu et al. \cite{xu2022connected} applied multi-agent reinforcement learning (MARL) for cooperative lane change control in convoy vehicles, while Fan et al. \cite{fan2023twin} optimized defensive decisions for escort vehicles using attention mechanisms. Notably, de Zarza et al. \cite{de2023decentralized} proposed the decentralized truck convoy framework O'Platoon based on Q-learning, which enables autonomous decision-making for individual vehicles through local information exchange, thereby enhancing system scalability while reducing communication dependency. These data-driven approaches offer higher flexibility but often face challenges related to training efficiency and generalization capabilities.

\subsection{LLM for Autonomous Driving} In recent years, large language models, represented by the GPT series and LLaMA series, have shown tremendous potential in a wide range of scenarios due to their strong contextual understanding and generative capabilities.

In the field of autonomous driving, researchers have begun to explore how the advantages of LLMs can be integrated with the needs of autonomous driving systems. Wen et al. were the first to apply the knowledge-driven capability of LLMs to autonomous vehicle decision-making, proposing the DiLu framework \cite{wen2023dilu}, which, however, was limited to single-agent decision-making. Jiang et al. introduced the KoMA framework \cite{jiang2024koma}, consisting of multi-agent interaction, multi-step planning, shared memory, and a ranking-based reflective module, which is a multi-agent framework. Similar frameworks include CoMAL \cite{yao2024comal} and CoDrivingLLM \cite{fang2024towards}.

With the rise of multimodal LLMs, some recent research has focused on how autonomous driving can be combined with visual LLMs, enabling visual LLMs to directly acquire information from images rather than from text descriptions provided by humans. Notable examples include Drivevlm \cite{tian2024drivevlm} and OmniDrive \cite{wang2024omnidrive}. These studies demonstrate the vast potential of LLM applications in autonomous driving, especially in complex dynamic environments such as convoy formation control.

\subsection{Formation Control} Efficient formation control is vital for multi-lane convoy operations, where dynamic traffic and unpredictable obstacles challenge traditional single-lane methods. Conventional approaches (e.g., leader–follower, virtual structure, and consensus-based strategies \cite{chen2023survey}) work well in regulated settings but often falter amid communication delays and rapid changes in convoy composition. Recent advances—such as Zhao et al.'s multi-platoon collision avoidance framework \cite{zhao2023multi} and their robust consensus method \cite{zhao2022consensus}, along with a hybrid automata approach for unified path planning and control \cite{huang2018path}—have improved formation stability in complex scenarios. Complementary reviews and developments in distributed control \cite{chu2024survey, yang2023distributed} further highlight these efforts. In contrast to these rule-based or data-driven models, our work employs a large language model to offer real-time, context-aware decision support, bridging the gap between traditional formation control and intelligent, flexible convoy management.

\section{METHODOLOGY}
In multi-lane convoy control, the central challenge is to achieve efficient inter-vehicle coordination and maintain stable formations in high-speed, dynamic traffic environments. Vehicles must not only execute routine driving tasks but also adaptively manage unexpected scenarios—such as obstacle avoidance, joining or leaving the convoy, and reconfiguring the formation—amid complex and variable road conditions. Traditional rule-based and reinforcement learning approaches, which are heavily data-dependent and exhibit limited generalization capabilities, often fall short in meeting the demands of real-time decision-making. Therefore, integrating the robust semantic understanding and reasoning capabilities of large language models into novel control strategies has emerged as a pivotal solution to this challenge.

\subsection{Overall Architecture}
\begin{figure*}[htbp]
    \centering
    \includegraphics[width=\textwidth]{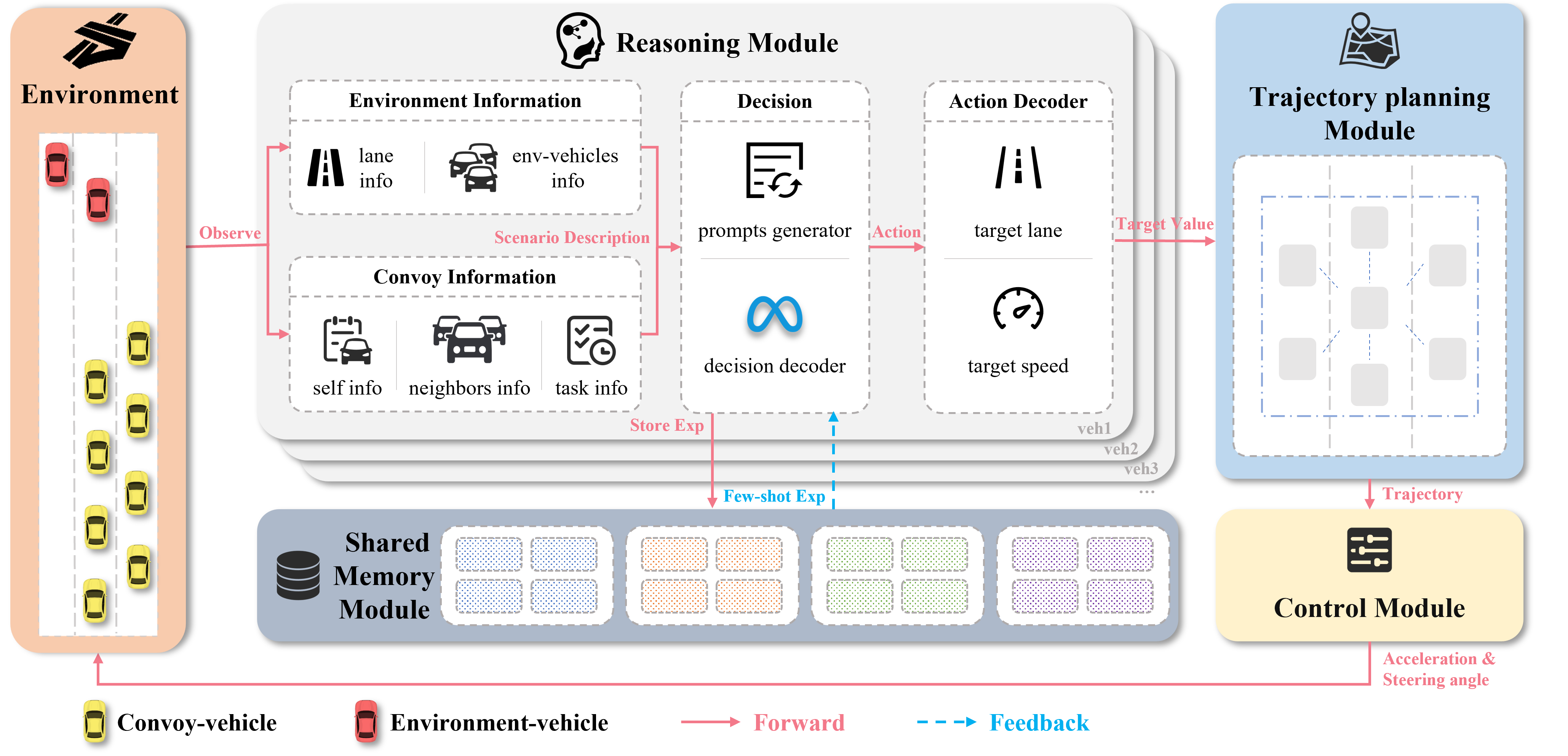}
    \caption{The overall framework of the multi-lane convoy formation control method. It contains a total of five modules: environment, reasoning, shared memory, trajectory planning, and control. The reasoning module obtains the perception results from the environment and generates the target lanes and target speeds of the vehicles, the trajectory planning module obtains the target values and generates the trajectories of each vehicle in the convoy, and finally the control module outputs the acceleration and steering angle commands, which are then applied to the environment.}
    \label{fig:overview}
\end{figure*}

The overall framework of the proposed method is shown in Fig. \ref{fig:overview}. This approach utilizes the independent large language model (LLM) of each vehicle for decision-making, while dynamically maintaining the overall structure of the convoy through a locally dynamic distributed graph. The framework comprises three core components: the reasoning module, the shared memory module, and the trajectory planning module. Each module performs its respective tasks and collaborates to ensure the convoy's flexibility, stability, and robustness.

In this framework, the reasoning module, centered around the LLM, is responsible for environment perception, high-level decision generation, and action decoding. The shared memory module uniformly stores and retrieves task experiences, facilitating cooperative learning among vehicles within the convoy. The trajectory planning module uses a locally dynamic distributed graph to enable dynamic formation adjustment and precise motion trajectory planning for the convoy. Additionally, the framework interacts with the external environment through the environment and control modules, further enhancing the real-time performance and adaptability in multi-lane scenarios.

\subsection{Reasoning Module}
The reasoning module is a critical component of the multi-lane convoy formation control system, designed to enable the entire process of vehicle reasoning from environment perception to high-level decision-making and action generation. This module employs chain reasoning, transforming environment and convoy information into high-level objectives, ensuring that connected autonomous vehicles can collaborate efficiently in dynamic traffic environments. The reasoning module consists of four key sub-module: the environment information, the convoy information, the decision, and the action decoder.

\textit{1) Environment Information:} The environment information sub-module is responsible for extracting the necessary perception data from the external environment, providing foundational support for subsequent reasoning and decision-making. This sub-module collects detailed road information, such as the number of lanes and the maximum allowed speed on the current road, and gathers key data about environment vehicles, including their lanes, precise positions, and real-time speeds. By structuring this data, the sub-module provides a clear description of the dynamic environment in which the vehicle operates for the reasoning module.

\textit{2) Convoy Information:} The convoy information sub-module consolidates the state information of the ego vehicle and neighboring vehicles, providing critical support for the reasoning module. The ego vehicle state information includes the current lane, precise position, and speed, while neighboring vehicle state information includes the lane, relative bearing, precise position, and real-time speed. Additionally, this sub-module retrieves the ego vehicle's task information, which includes four types of tasks: avoiding obstacles, joining the convoy, leaving the convoy, and switching to an escort formation.
\begin{figure*}[htbp]
    \centering
    \includegraphics[width=\textwidth]{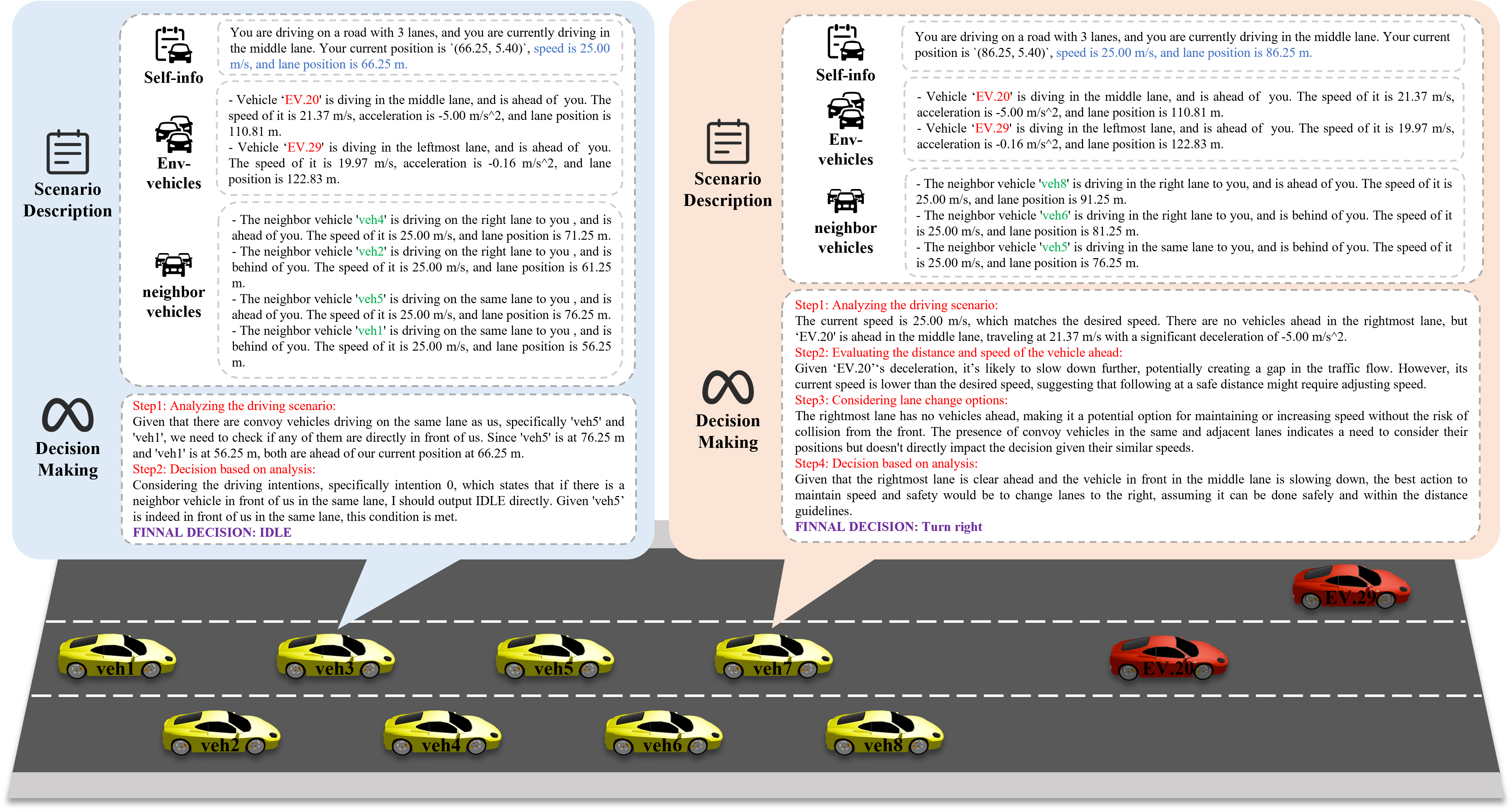}
    \caption{A case of the Reasoning module process. This simple obstacle avoidance scenario illustrates how the reasoning module collects information from the ego vehicle, environment vehicles, and neighbors, then generates a scene description for decision-making by the large model. In the figure, veh7 changes lanes to the right due to a slow vehicle ahead, while veh3 outputs an IDLE decision to follow the neighboring vehicle in the same lane.}
\end{figure*}

\textit{3) Decision:} As the core of the reasoning module, the decision sub-module generates high-level decisions for the convoy vehicles to address the dynamic and complex multi-lane traffic environment. This sub-module consists of the \textbf{prompts generator} and the \textbf{decision decoder}, which collaborate to complete the full process from scene analysis to action planning.

The prompts generator constructs scene descriptions based on environment information and convoy information, including road conditions, environment vehicle states, neighboring vehicle information, and task objectives. These descriptions are then input into the LLM in a unified format to provide the correct context for reasoning. The decision decoder, using Few-shot learning, converts scene descriptions into specific decisions, including IDLE, left lane change, right lane change, accelerate, and decelerate, thereby meeting the basic needs of complex and dynamic traffic scenarios.

Furthermore, the decision sub-module closely interacts with the shared memory module by retrieving similar historical experiences to provide the LLM with prior knowledge, improving the accuracy and flexibility of decisions. 

\textit{4) Action Decoder:} The action decoder is responsible for converting the high-level action commands output by the decision sub-module into specific low-level control targets. Specifically, the action decoder generates the target lane and target speed, and passes these target values to the trajectory planning module for execution. 
\subsection{Shared Memory Module}
The shared memory module is a key design in the system to support vehicle task learning and knowledge transfer, which is used to realize collaborative learning and efficient decision making among convoy vehicles in different task scenarios. For the complexity of multi-lane convoy formation control, we designed the following four main task scenarios: avoiding obstacles in the convoy, joining the convoy, leaving the convoy, and switching to escort formation, which means that convoy vehicles may be performing one of these four tasks. In order to effectively manage these task experiences, the module is designed with four separate areas corresponding to each of these four tasks. The shared memory module contains the following two main functions.

\textit{1) Experience Storage Mechanism:} The shared memory module adopts a task-oriented experience pool (TEP) architecture, where all experiences generated by vehicles performing the same task are stored in the corresponding task area, rather than being stored solely within individual vehicles. For example, all experiences generated by vehicles performing the "joining the convoy" task are collectively stored in the "joining the convoy" area. Similarly, experiences for the other tasks are categorized and stored in their respective areas.

\textit{2) Decision Support Function:} The shared memory module provides strong support for the large language model in the reasoning module. When a vehicle needs to make a real-time decision, the LLM retrieves several experiences from the shared memory module that are similar to the current scenario and uses them as input to assist in the reasoning process. By referencing similar experiences, the LLM can more accurately understand the characteristics of the current task and generate better decision outcomes. For example, when a vehicle needs to join the convoy, the reasoning module can extract similar historical experiences from the "joining the convoy" task area to help the vehicle complete the task more smoothly.

\subsection{Trajectory Planning Module}  
The trajectory planning module, as the core of the multi-lane convoy control system, achieves collaborative motion control of vehicles through a locally dynamic distributed graph, based on the target lane and speed output by the reasoning module. As shown in Fig. \ref{fig:local_graph}, the system adopts an interlaced formation structure and defines six types of neighboring nodes to enhance flexibility:

\begin{figure}[htp]
    \centering
    \includegraphics[width=\linewidth]{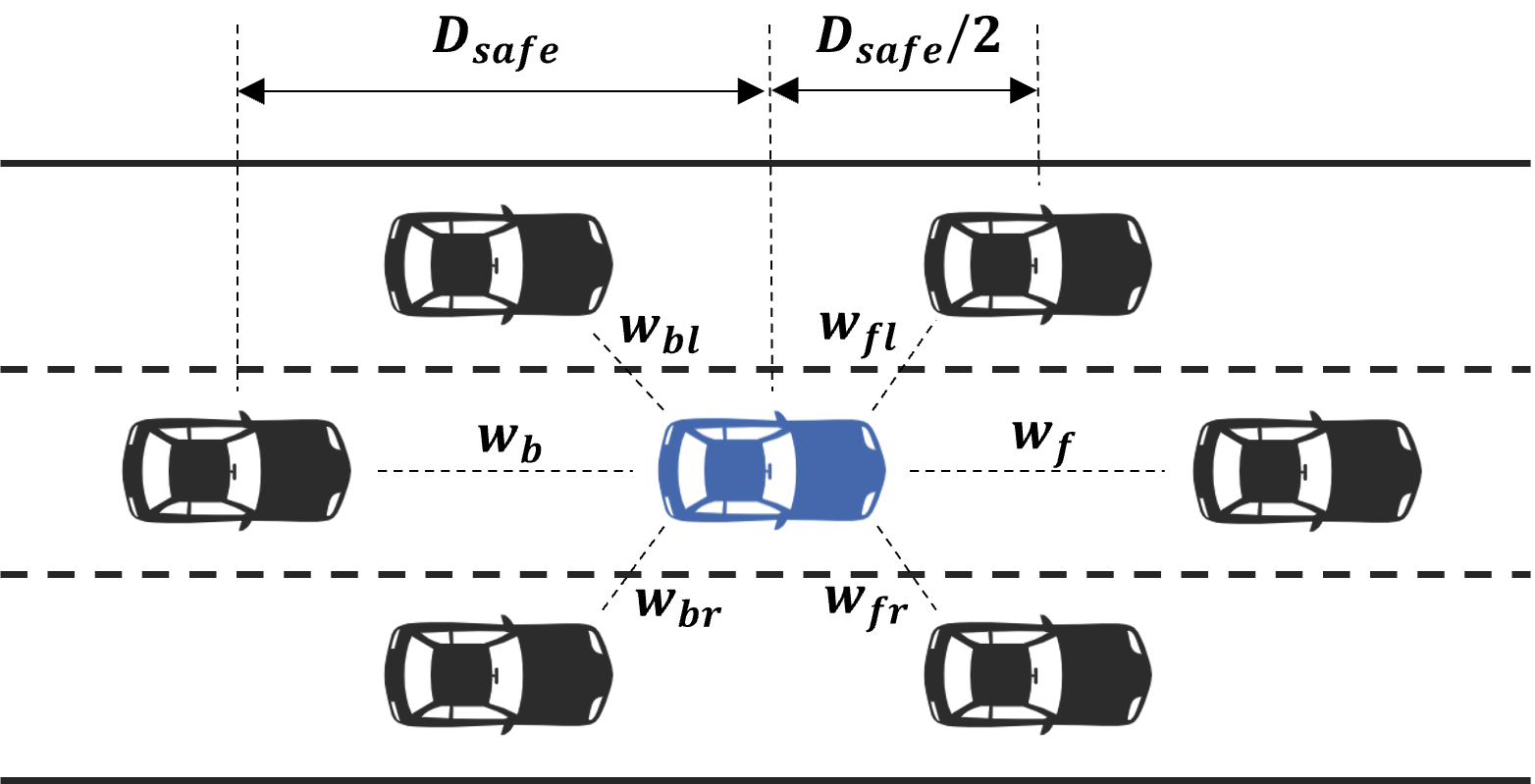}
    \caption{Interlaced formation and neighboring nodes diagram}
    \label{fig:local_graph}
\end{figure}

\textbf{Interlaced Formation}: Compared to a parallel formation, the interlaced formation offers better flexibility, providing more lane-change space for vehicles in adjacent lanes, significantly reducing the risk of merging conflicts.

\textbf{Neighboring Nodes}: The nearest vehicles in the forward and backward directions within the ego vehicle’s communication range on any lane are defined as neighboring nodes, including:
\begin{itemize}
    \item Forward/backward vehicle in the same lane ($N_f$, $N_b$)
    \item Forward vehicle in the left/right adjacent lane ($N_{fl}$, $N_{fr}$)
    \item Backward vehicle in the left/right adjacent lane ($N_{bl}$, $N_{br}$)
\end{itemize}

The neighboring set $\mathcal{N}=\{N_{fr},N_{br},N_f,N_b,N_{fl},N_{bl}\}$ is dynamically updated, with missing nodes filled with zero values.

Based on the local fully connected graph topology, a distributed control law is designed:
\begin{equation}
\begin{bmatrix}
\dot{x} \\ \dot{y}
\end{bmatrix} =
\left[
\begin{array}{c}
\sum\limits_{n \in \mathcal{N}} w_n \cdot ((x_n-x_{ego}) - d_{desired}) \\
w_y \cdot ( y_{target\_lane} - y_{ego})
\end{array}
\right] + 
\begin{bmatrix}
v_{target\_x} \\ 0
\end{bmatrix}
\end{equation}

Where $n$ is a neighboring node, $w_n$ is the weight coefficient between the ego vehicle and the neighboring node, $x_n$ is the longitudinal coordinate of the neighboring node, $x_{ego}$ is the longitudinal coordinate of the ego vehicle, $d_{desired}$ is the desired distance between the ego vehicle and the neighboring node. If the ego vehicle and the neighboring node are on the same lane, then $d_{desired}=D_{safe}$; otherwise, $d_{desired}=D_{safe}/2$. $v_{target\_x}$ is the target speed output by the reasoning module, $y_{ego}$ is the ego vehicle’s current lateral position, and $y_{target\_lane}$ is the lateral position of the target lane centerline as output by the reasoning module.

The speed of vehicles on the same lane within the convoy should also remain consistent to avoid collisions. When the first vehicle in a lane accelerates or decelerates, the following vehicles will select the IDLE action to follow, resulting in a speed difference with the front vehicle. To maintain speed consistency within the same lane, a speed coordination strategy is designed:
\begin{equation}
    \dot{v}=w_{v} \cdot (v_f-v_{ego})
\end{equation}

Where $w_{v}$ is the weight coefficient, $v_f$ is the speed of the forward neighboring node in the same lane, and $v_{ego}$ is the ego vehicle’s current speed.

The control module in this work utilizes the validated controller from \cite{gao2019multi}, ensuring stable trajectory tracking, while our primary contribution lies in the LLM-driven high-level decision-making framework.

\section{EXPERIMENTS}  
To evaluate the effectiveness and robustness of the proposed multi-lane convoy control method based on LLM, we conducted extensive experiments in the SUMO (Simulation of Urban MObility) simulation environment. Four typical scenarios were designed: avoiding obstacles, joining the convoy, leaving the convoy, and switching to a escort formation. All experiments were conducted on a 1000-meter long highway with three lanes, where the convoy consisted of 8 CAVs. These scenarios were designed to comprehensively assess the flexibility, stability, and adaptability of the proposed method in dynamic and complex traffic environments.

In this study, we used LLaMA-3.3 as the base large language model, and conducted all experiments under a 3-shot learning paradigm. In the experimental setup, we defined key parameters such as the weight coefficients, desired speed of the convoy vehicles ($v_{desired}$), acceleration ($acc$), and deceleration ($dec$), as shown in Table \ref{tab:experiment_parameters}.

\begin{table}[H]
    \centering
    \caption{Experiment Parameters}
    \label{tab:experiment_parameters}
    \small
    \renewcommand{\arraystretch}{1.2}
    \begin{tabular}{c|c|c|c}
        \hline
        \textbf{Parameter} & \textbf{Value} & \textbf{Parameter} & \textbf{Value} \\ \hline
        \( D_{\text{safe}}(m) \) & 10 & \( w_{\text{bl}} \) & 0.1 \\ \hline
        \( w_{\text{fr}} \) & 1.0 & \( w_{\text{y}} \) & 1.8 \\ \hline
        \( w_{\text{br}} \) & 0.1 & \( w_{\text{v}} \) & 0.5 \\ \hline
        \( w_{\text{f}} \) & 2.0 & \( v_{\text{desired}}(m/s) \) & 25 \\ \hline
        \( w_{\text{b}} \) & 1.0 & \( acc(m/s^2) \) & 1.0 \\ \hline
        \( w_{\text{fl}} \) & 1.0 & \( dec(m/s^2) \) & 2.0 \\ \hline
    \end{tabular}
\end{table}
\subsection{Avoiding Obstacles}  
\begin{figure}[htbp]
    \centering  
    \begin{subfigure}[b]{\linewidth}  
        \includegraphics[width=\linewidth]{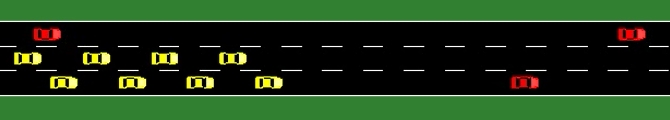}  
        \caption{20 environment vehicles}  
        \label{fig:subfig1}  
    \end{subfigure}  

    \begin{subfigure}[b]{\linewidth}  
        \includegraphics[width=\linewidth]{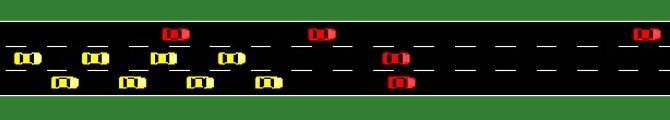}  
        \caption{30 environment vehicles}  
        \label{fig:subfig2}  
    \end{subfigure}  

    \begin{subfigure}[b]{\linewidth}  
        \includegraphics[width=\linewidth]{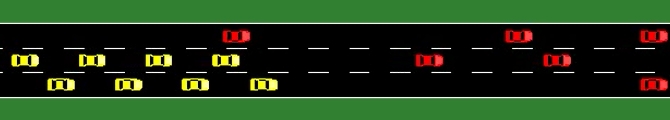}  
        \caption{40 environment vehicles}  
        \label{fig:subfig3}  
    \end{subfigure}  
    \caption{Avoiding obstacles scenario}  
    \label{fig:scene1}  
\end{figure}

In the obstacle avoidance scenario, the convoy must dynamically maneuver to avoid obstacles during its journey, and, when necessary, adapt its formation to maintain both safety and efficiency. At the beginning of the experiment, a certain number of environment vehicles were randomly generated within the first 700 meters of the highway. We set three different traffic densities, corresponding to 20, 30, and 40 environment vehicles. The speed of the environment vehicles was randomly initialized between 15 m/s and 30 m/s. The experimental scenario is shown in Fig. \ref{fig:scene1}, where yellow vehicles represent convoy vehicles and red vehicles represent environment vehicles. 

Success was defined as all convoy vehicles completing the 1000-meter journey collision-free. To evaluate the performance of the experiments, we used two metrics: success rate and average speed, and we conducted 50 independent experiments under each traffic density setup using different random seeds.

\begin{table}[H]  
    \centering  
    \caption{Success rate under different traffic densities}  
    \label{table:SR}  
    \small  
    \renewcommand{\arraystretch}{1.5}
    \begin{tabular}{c|c|c|c}  
        \hline  
        \textbf{Environment vehicle number} & \text{20} & \text{30} & \text{40} \\ \hline  
        \textbf{Success rate} & 76\% & 72\% & 64\% \\ \hline  
    \end{tabular}  
\end{table}  

\begin{figure}[htbp]  
    \centering  
    \begin{subfigure}[b]{\linewidth}  
        \includegraphics[width=\linewidth]{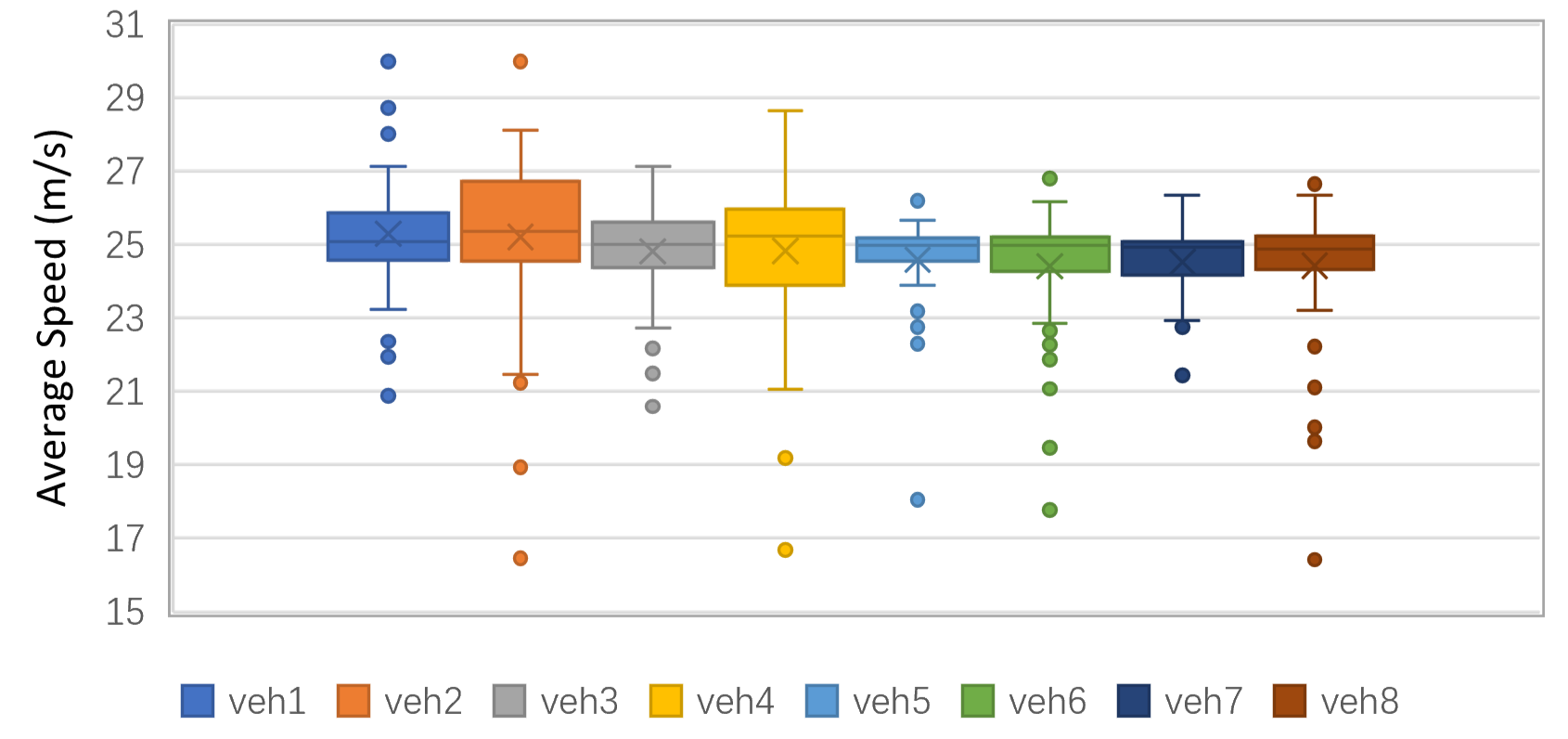}  
        \caption{20 environment vehicles}  
        \label{fig:subfig1_AS}  
    \end{subfigure}  
    \begin{subfigure}[b]{\linewidth}  
        \includegraphics[width=\linewidth]{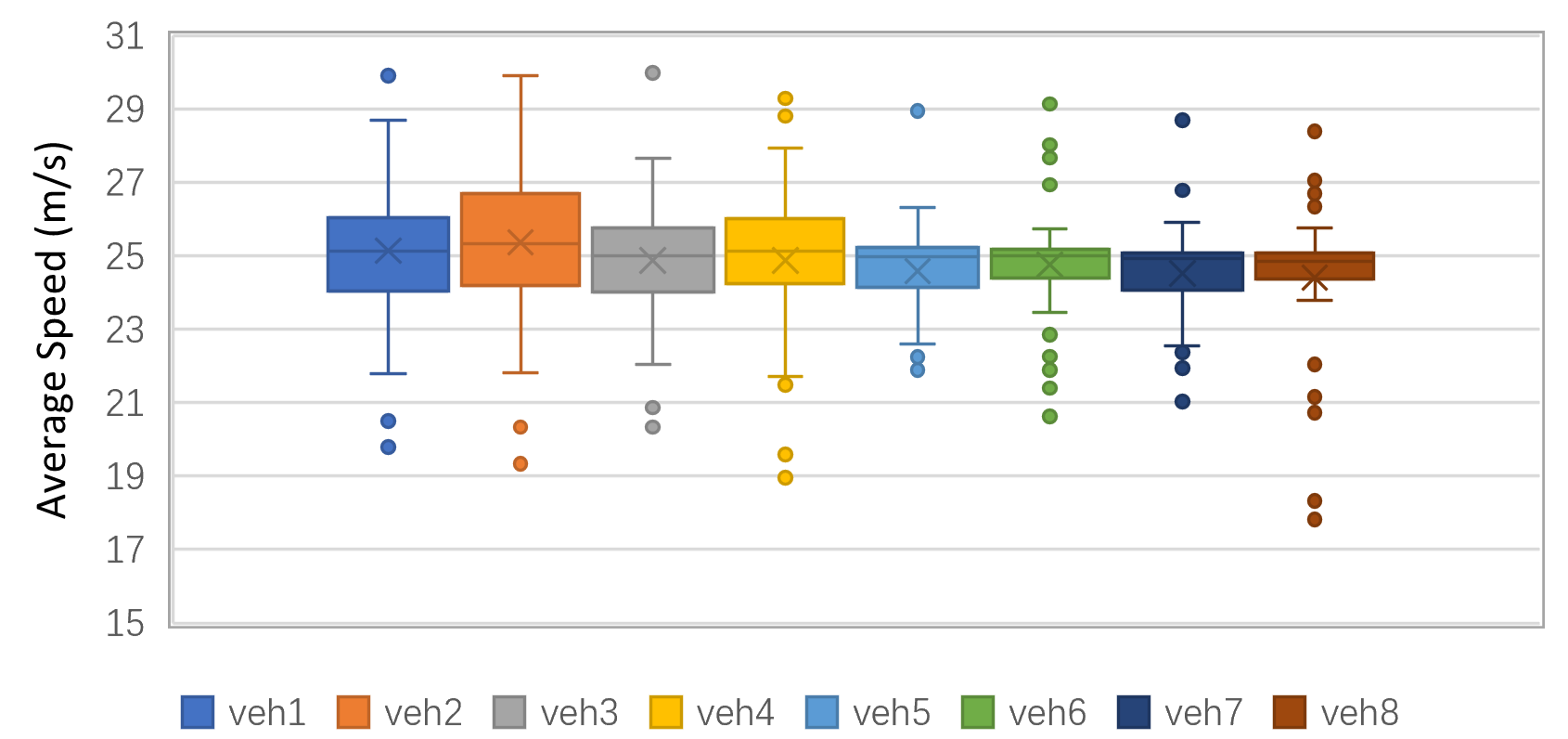}  
        \caption{30 environment vehicles}  
        \label{fig:subfig2_AS}  
    \end{subfigure}  
    \begin{subfigure}[b]{\linewidth}  
        \includegraphics[width=\linewidth]{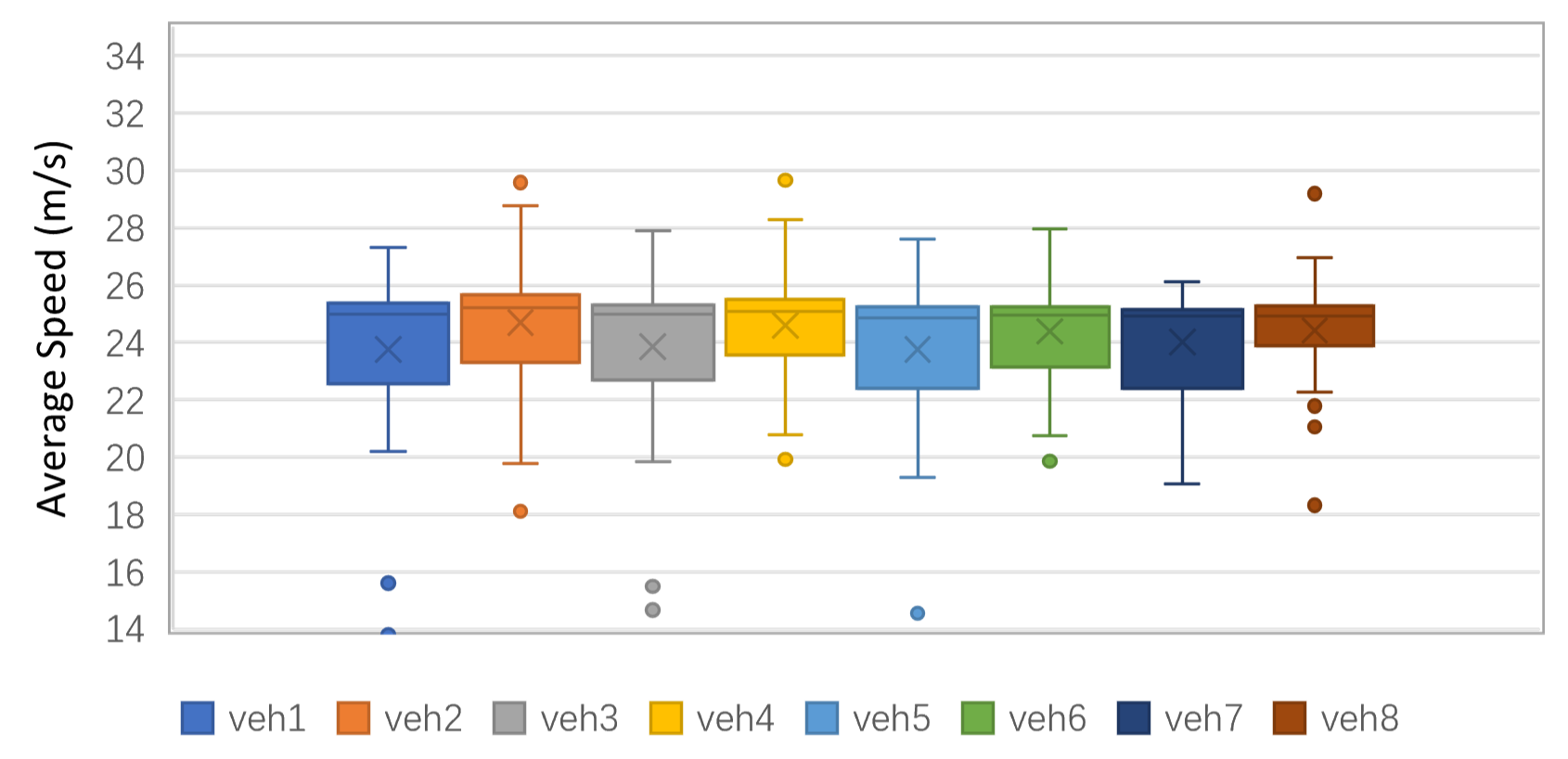}  
        \caption{40 environment vehicles}  
        \label{fig:subfig3_AS}  
    \end{subfigure}  
    \caption{Average speed distribution of convoy under different traffic densities}  
    \label{fig:average_speed}  
\end{figure}  

The success rates for different traffic densities are shown in Table \ref{table:SR}. The proposed method achieved an 76\% success rate in the low-density scenario (20 environment vehicles), 72\% success rate in the medium-density scenario (30 environment vehicles), and 64\% success rate in the high-density scenario (40 environment vehicles). As shown in Fig. \ref{fig:average_speed}, the average speed of the convoy remained stable at around 25 m/s in both the low and medium-density scenarios. However, in the high-density scenario, the average speed concentrated between 23 m/s and 25 m/s. This was due to the increased number of slow-moving environment vehicles in the high-density scenario, which made it difficult for the convoy to overtake and maintain the desired speed, leading the convoy to reduce speed and follow the environment vehicles. The experimental results demonstrate that the proposed convoy control method effectively balances safety and efficiency in a dynamic traffic environment.

\subsection{Joining The Convoy}  
\begin{figure}[htbp]  
    \centering  
    \begin{subfigure}[b]{\linewidth}  
        \includegraphics[width=\linewidth]{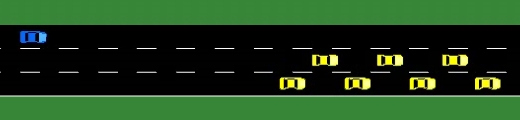}  
        \caption{Initial scenario}  
    \end{subfigure}  
    \begin{subfigure}[b]{\linewidth}  
        \includegraphics[width=\linewidth]{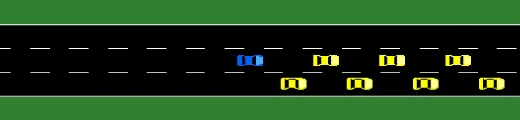}  
        \caption{Successful scenario}  
    \end{subfigure}  
    \caption{Joining the convoy scenario}  
    \label{fig:scene2}  
\end{figure}

The scenario is shown in Fig. \ref{fig:scene2}, where we select a vehicle from the convoy (highlighted in blue) and position it 50 meters behind the main convoy. This setup is designed to evaluate the ability of the proposed LLM-based method to dynamically integrate a vehicle into the convoy under real-time traffic conditions.  

In this experiment scenario, as well as in the next two, we use speed and position error (PE) as evaluation metrics. The calculation of PE is as follows:  
\begin{equation}  
    PE = \sum\limits_{n \in \mathcal{N}}{|(x_n - x_{ego}) - d_{desired}|}  
\end{equation}  

\begin{figure}[htbp]  
    \centering  
    \begin{subfigure}[b]{\linewidth}  
        \includegraphics[width=\linewidth]{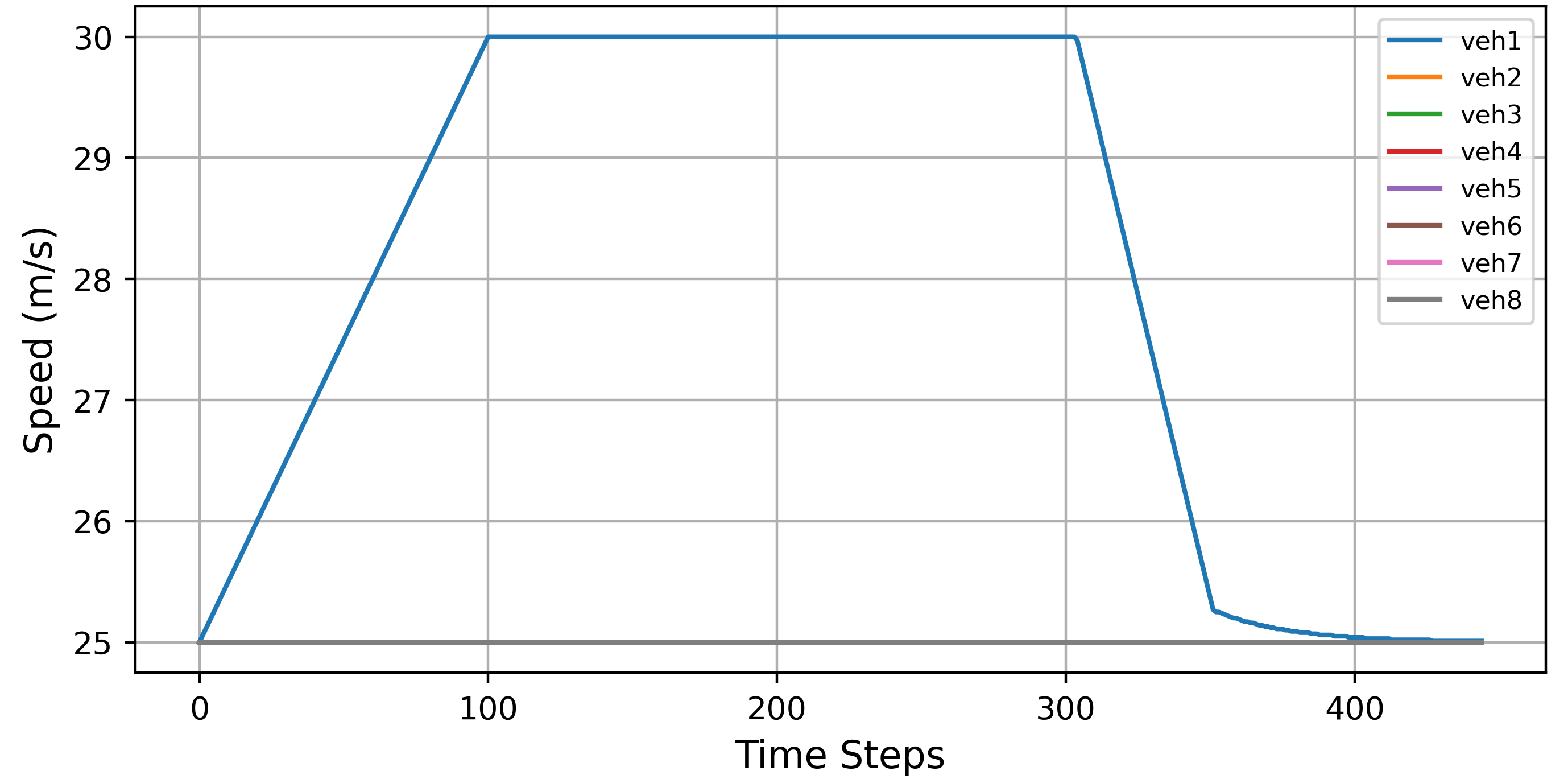}  
        \caption{Speed}  
    \end{subfigure}  
    \begin{subfigure}[b]{\linewidth}  
        \includegraphics[width=\linewidth]{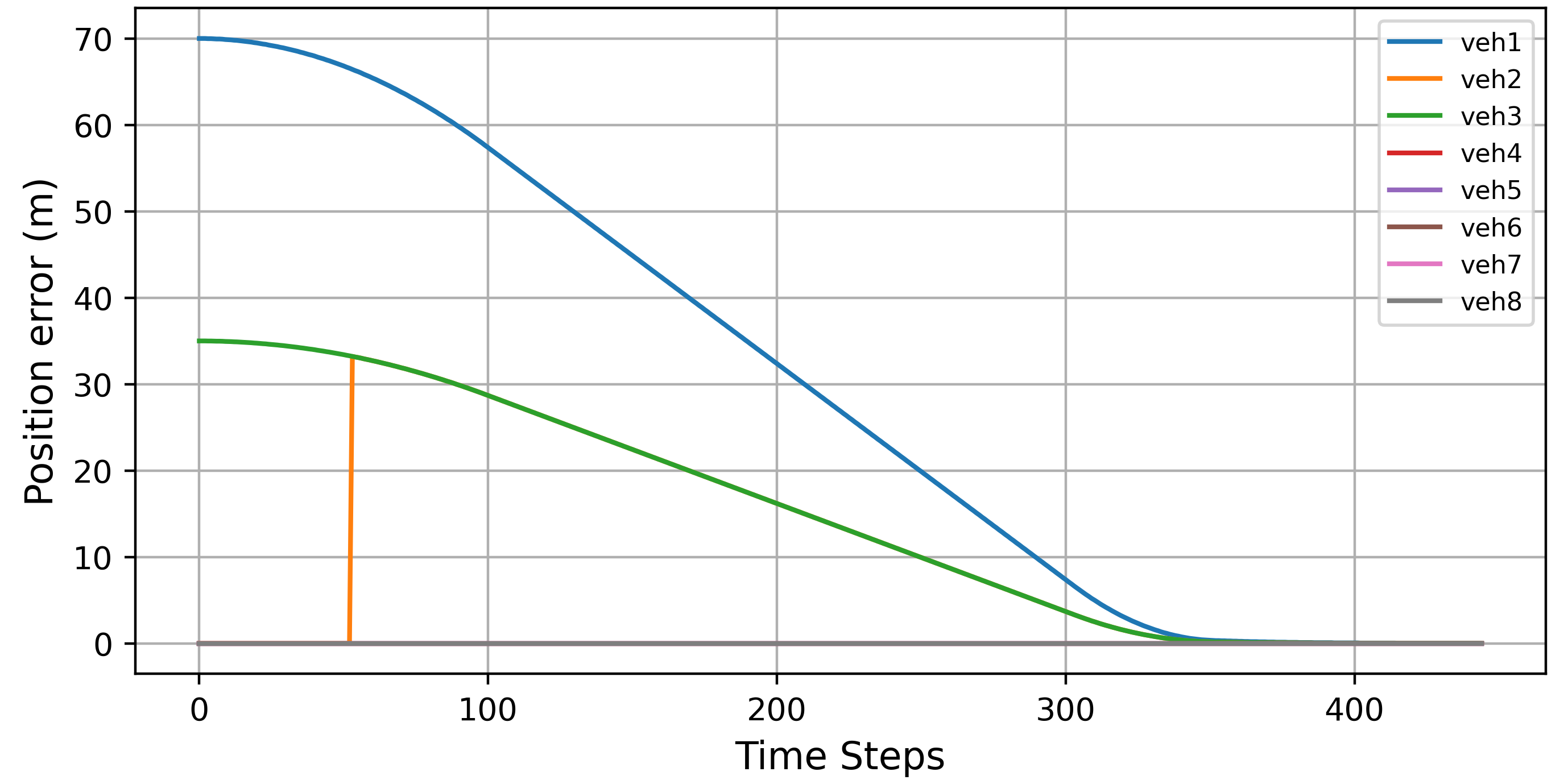}  
        \caption{Position error}  
    \end{subfigure}  
    \caption{The experiment results of joining the convoy scenario}  
    \label{fig:experiment_result2}  
\end{figure}  

The variations in speed and position error of each vehicle are shown in Fig. \ref{fig:experiment_result2}. The blue vehicle (veh1) first chooses to change lanes to the middle lane while simultaneously accelerating. It completes the lane change in about 50 time steps and continues to accelerate to the speed limit of 30 m/s. As it approaches the convoy, the vehicle gradually decelerates to align with the convoy and finally slows down to the desired speed after successfully joining the convoy.  

\subsection{Leaving The Convoy}  
\begin{figure}[htbp]  
    \centering  
    \begin{subfigure}[b]{\linewidth}  
        \includegraphics[width=\linewidth]{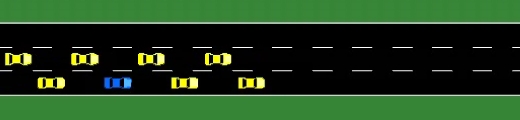}  
        \caption{Initial scenario}  
    \end{subfigure}  
    \begin{subfigure}[b]{\linewidth}  
        \includegraphics[width=\linewidth]{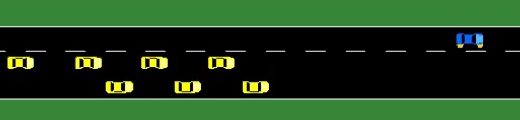}  
        \caption{Successful scenario}  
    \end{subfigure}  
    \caption{Leaving the convoy scenario}  
    \label{fig:scene3}  
\end{figure}

The scenario is shown in Fig. \ref{fig:scene3}, where a vehicle (highlighted in blue) in the convoy needs to leave the convoy for some reason. The success criterion is that the vehicle must exit the communication range of all vehicles in the convoy. The LLM-based decision module guides the vehicle to change lanes to the leftmost lane and then either accelerate or decelerate to distance itself from the convoy.

\begin{figure}[htbp]  
    \centering  
    \begin{subfigure}[b]{\linewidth}  
        \includegraphics[width=\linewidth]{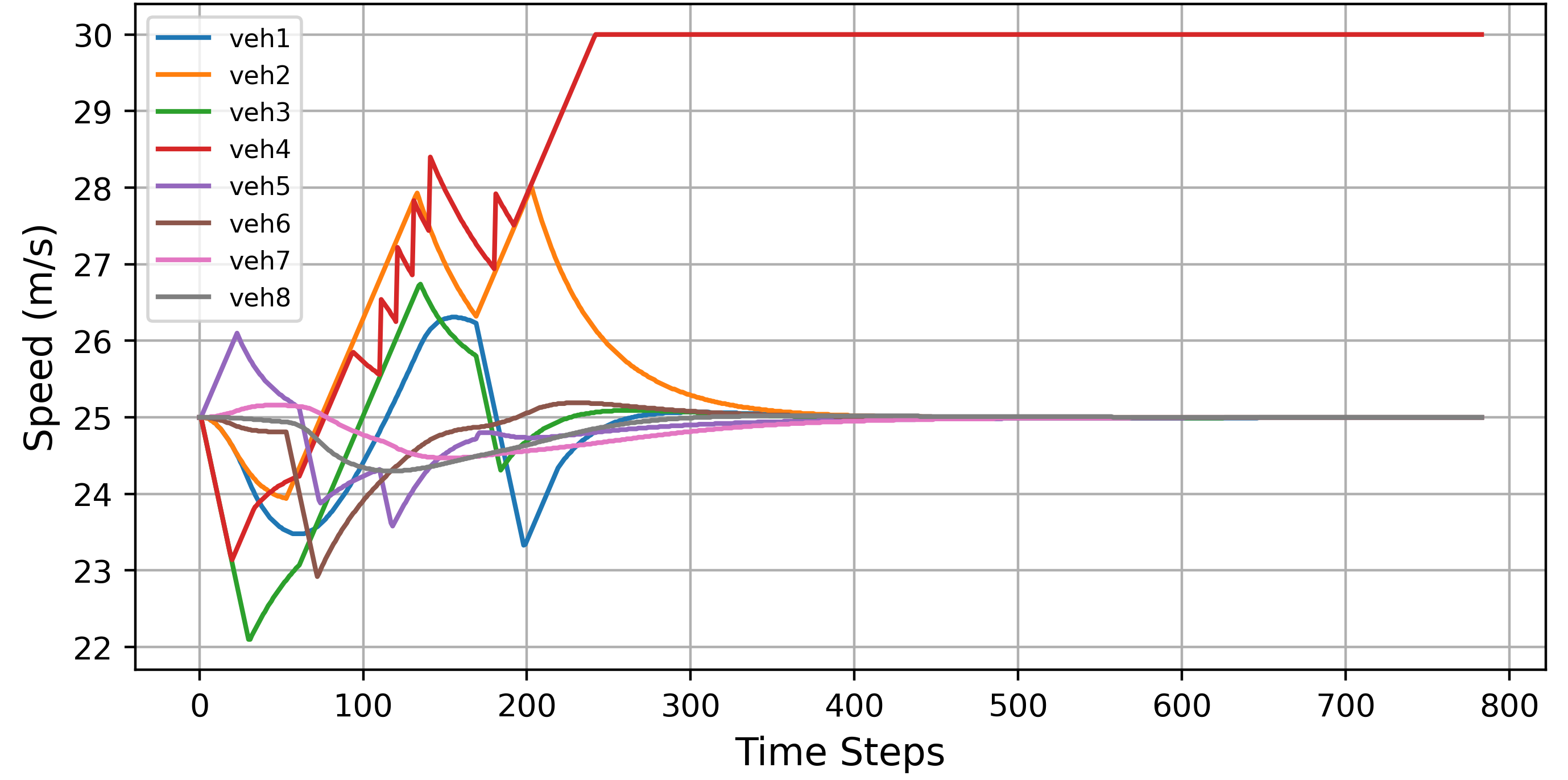}  
        \caption{Speed}  
        \label{fig:experiment_result3_speed}  
    \end{subfigure}  
    \begin{subfigure}[b]{\linewidth}  
        \includegraphics[width=\linewidth]{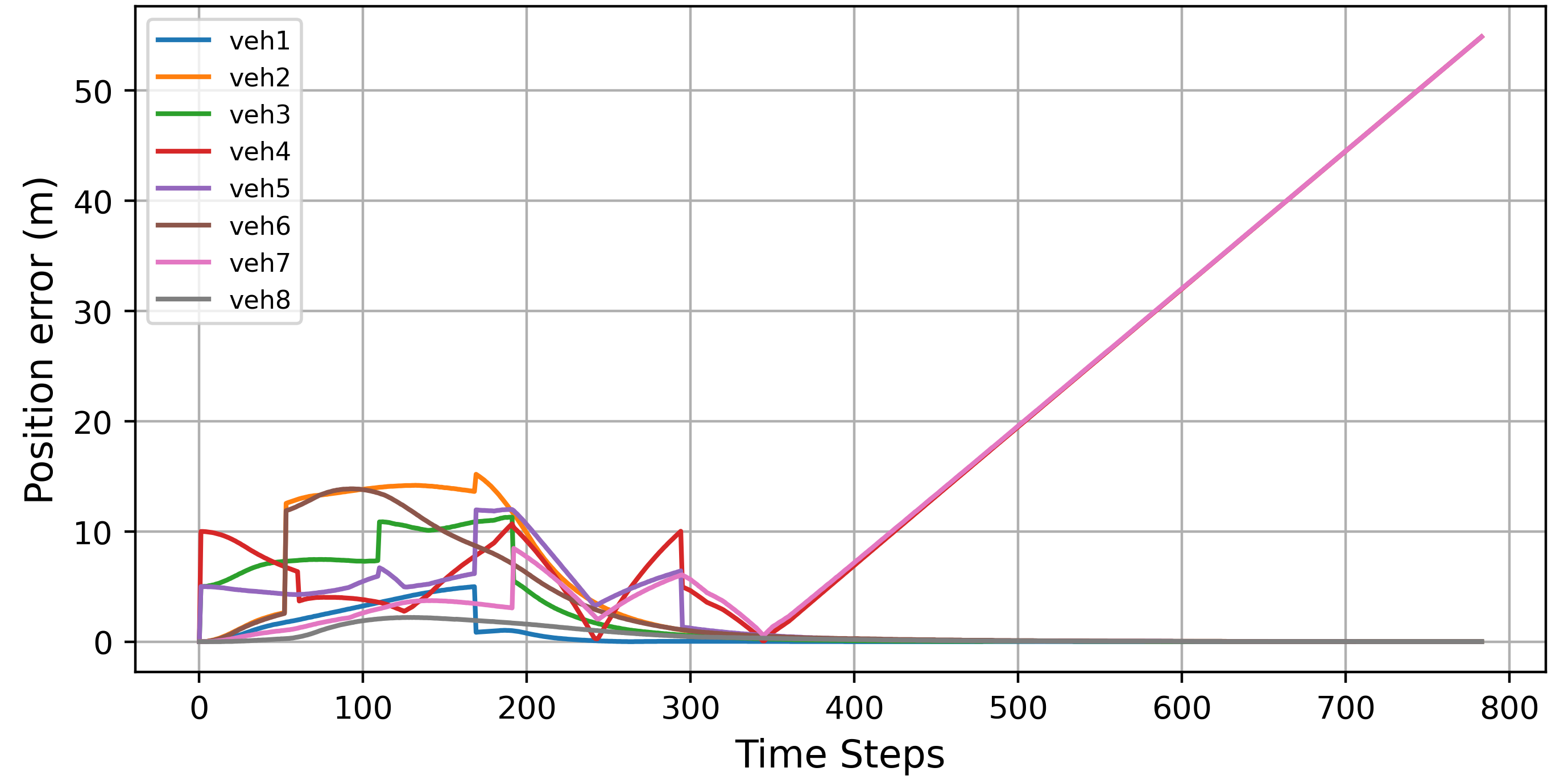}  
        \caption{Position error}  
        \label{fig:experiment_result3_PE}  
    \end{subfigure}  
    \caption{The experiment results of leaving the convoy scenario}  
    \label{fig:experiment_result3}  
\end{figure}

The variations in speed and position error of each vehicle in this scenario are shown in Fig. \ref{fig:experiment_result3}. The speed of the blue vehicle (veh4) increases gradually from the desired speed of 25 m/s to 30 m/s. Since veh4 needs to change lanes from the rightmost lane to the leftmost lane, the speed and position error of all vehicles fluctuate during the lane change. Once veh4 completes the lane change, other vehicles gradually recover to the desired speed, and the position error approaches 0. The convoy gradually stabilizes, while veh4 continues to accelerate away from the convoy, ultimately successfully leaving the convoy.

\subsection{Switching to Escort Formation}  
\begin{figure}[htbp]  
    \centering  
    \begin{subfigure}[b]{\linewidth}  
        \includegraphics[width=\linewidth]{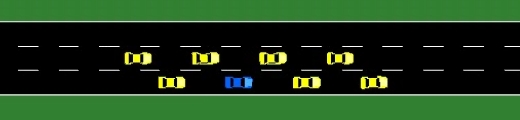}  
        \caption{Initial scenario}  
    \end{subfigure}  
    \begin{subfigure}[b]{\linewidth}  
        \includegraphics[width=\linewidth]{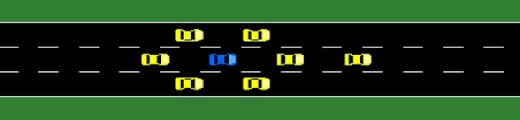}  
        \caption{Successful scenario}  
    \end{subfigure}  
    \caption{Switching to escort convoy scenario}  
    \label{fig:scene4}  
\end{figure}

This scenario is shown in Fig. \ref{fig:scene4}, involves the convoy switching to an escort formation to protect the blue vehicle(veh4).

\begin{figure}[htbp]  
    \centering  
    \begin{subfigure}[b]{\linewidth}  
        \includegraphics[width=\linewidth]{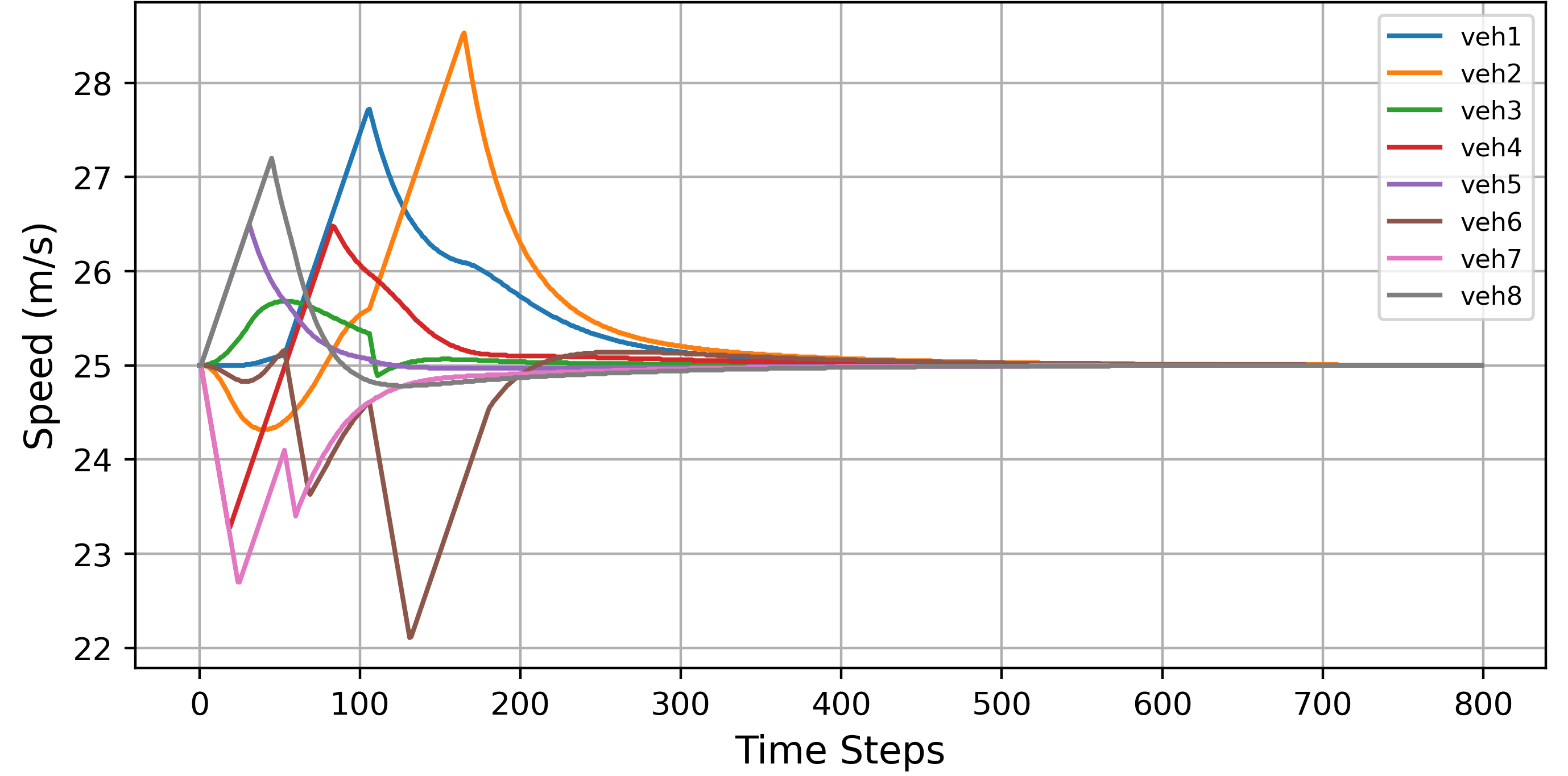}  
        \caption{Speed}  
    \end{subfigure}  
    \begin{subfigure}[b]{\linewidth}  
        \includegraphics[width=\linewidth]{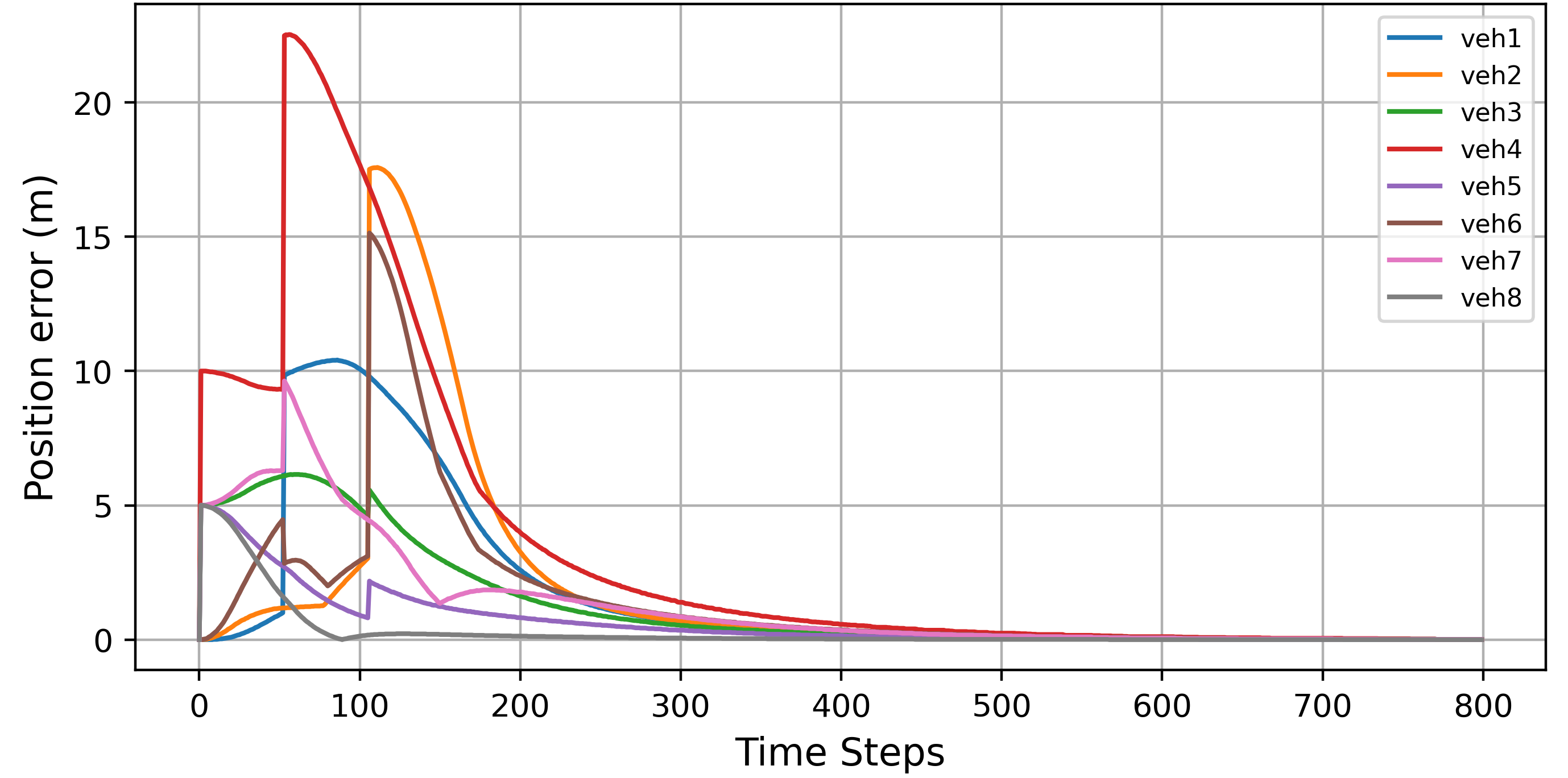}  
        \caption{Position error}  
    \end{subfigure}  
    \caption{The experimental result switching to escort formation scenario}  
    \label{fig:experiment_result4}  
\end{figure}

The variations in speed and position error of each vehicle in this scenario are shown in Fig. \ref{fig:experiment_result4}. During the formation switch, the vehicles either accelerate or decelerate due to lane changes or the influence of other vehicles changing lanes. As a result, the speed and position errors of all vehicles fluctuate. After approximately 400 time steps, the formation switch is completed, and the speeds of all vehicles gradually approach the desired speed, with position errors reducing to 0.

\section{CONCLUSIONS}  
This paper proposes a multi-lane convoy formation control method based on LLMs. By combining the reasoning capability of LLMs with a local dynamic distributed graph control strategy, we have enabled the convoy to maintain a stable formation while exhibiting flexible task execution capabilities.  The experimental results demonstrate that the proposed method performs well across four typical scenarios (avoiding obstacles, joining the convoy, leaving the convoy, and switching to escort formation), showing notable advantages in adaptability and robustness.

Future work will focus on optimizing LLM reasoning efficiency, integrating multi-modal inputs (e.g., vision and radar), and real-world deployment.  Our approach advances collaborative control for intelligent connected vehicles, offering practical potential in traffic management and logistics.

\addtolength{\textheight}{-12cm}   % This command serves to balance the column lengths

%%%%%%%%%%%%%%%%%%%%%%%%%%%%%%%%%%%%%%%%%%%%%%%%%%%%%%%%%%%%%%%%%%%%%%%%%%%%%%%%

%%%%%%%%%%%%%%%%%%%%%%%%%%%%%%%%%%%%%%%%%%%%%%%%%%%%%%%%%%%%%%%%%%%%%%%%%%%%%%%%

%%%%%%%%%%%%%%%%%%%%%%%%%%%%%%%%%%%%%%%%%%%%%%%%%%%%%%%%%%%%%%%%%%%%%%%%%%%%%%%%

% \begin{thebibliography}{99}
% \begin{refcontext}[sorting = none]
% \printbibliography
\bibliography{ref}
% \end{refcontext}
% \end{thebibliography}

\end{document}